# 4D Real-Time GRASP MRI at Sub-Second Temporal Resolution

Li Feng

Biomedical Engineering and Imaging Institute and Department of Radiology, Icahn School of Medicine at Mount Sinai, New York, NY, United States

**Grant Support:** NIH R01EB030549, R01EB031083, R21EB032917

Address correspondence to:

    Li Feng, PhD

    Biomedical Engineering and Imaging Institute

    Department of Radiology

    Icahn School of Medicine at Mount Sinai

    1470 Madison Ave

    New York, NY, USA 10029

    Email: lifeng.mri@gmail.com

## Abstract

Intra-frame motion blurring, as a major challenge in free-breathing dynamic MRI, can be reduced if high temporal resolution can be achieved. To address this challenge, this work proposes a highly-accelerated 4D (3D+time) real-time MRI framework with sub-second temporal resolution combining standard stack-of-stars golden-angle radial sampling and tailored GRASP-Pro (Golden-angle RAdial Sparse Parallel) reconstruction. Specifically, 4D real-time MRI acquisition is performed continuously without motion gating or sorting. The k-space centers in stack-of-stars radial data are organized to guide estimation of a temporal basis, with which GRASP-Pro reconstruction is employed to enforce joint low-rank subspace and sparsity constraints. This new basis estimation strategy is the new feature proposed for subspace-based reconstruction in this work to achieve high temporal resolution (e.g., sub-second/3D volume). It does not require sequence modification to acquire additional navigation data, is compatible with commercially available stack-of-stars sequences, and does not need an intermediate reconstruction step. The proposed 4D real-time MRI approach was tested in abdominal motion phantom, free-breathing abdominal MRI, and dynamic contrast-enhanced MRI (DCE-MRI). Our results have shown that GRASP-Pro reconstruction with the new basis estimation strategy enabled highly-accelerated 4D dynamic imaging at sub-second temporal resolution (with 5 spokes or less for each dynamic frame per image slice) for both free-breathing non-DCE-MRI and DCE-MRI. In the abdominal phantom, better image quality with lower RMSE and higher SSIM was achieved using GRASP-Pro compared to standard GRASP reconstruction. With the ability to acquire each 3D image in less than one second, intra-frame respiratory blurring can be intrinsically reduced for body applications with our approach, which also eliminates the need for motion detection and motion compensation.

## Introduction

High-performance volumetric MRI (3D MRI) in presence of respiratory motion has been a challenging research topic that attracts substantial interest and attention. Due to the slow acquisition speed in MRI, state-of-the-art free-breathing 3D imaging methods typically rely on triggering, gating, registration-based motion correction, motion-weighted reconstruction, or motion-resolved reconstruction to reduce respiration-induced motion effect (1–5) (e.g., blurring and/or ghosting) within an image (referred to as intra-frame blurring in dynamic imaging). To date, these approaches still suffer from different challenges and limitations at varying degrees, such as reduced imaging efficiency, compromised image quality under irregular breathing, and sensitivity to motion drifts. In the meantime, these approaches also require accurate information about the underlying respiratory motion pattern (typically obtained through a process referred to as motion detection) for motion compensation. Inaccurate motion detection can lead to poor motion compensation, yielding residual motion artifacts. Currently, motion detection is commonly performed with external motion sensors (e.g., a respiratory bellow or other new devices such as the pilot tone (6,7)) or with self-navigation to extract a respiratory motion signal (8–12). For dynamic contrast-enhanced MRI (DCE-MRI), respiratory motion detection can be much more challenging due to the change of image contrast along time; and therefore, the detected motion signal can be contaminated by the contrast variation (4,13). This potentially reduces the accuracy and robustness of motion detection in free-breathing DCE-MRI, which constitutes a major restriction for its wide clinical translation.

Stack-of-stars golden-angle radial sampling has emerged as a promising hybrid acquisition trajectory (in-plane radial sampling + through-plane Cartesian sampling) for free-breathing MRI and DCE-MRI (14,15). In radial imaging, motion artifacts typically spread through the entire field of view (FOV) as blurring, and adequate image quality has been demonstrated even with mild motion blurring (16–18). Besides, radial sampling provides good incoherence that can be exploited with sparse reconstruction techniques for imaging acceleration (19,20), and the golden-angle rotation scheme enables simple push-bottom data acquisition with flexibility to reconstruct dynamic images with arbitrary temporal resolution (14,21,22). Combining stack-of-stars golden-angle radial sampling with multicoil compressed sensing reconstruction, a method called GRASP (Golden-angle RAdial Sparse Parallel imaging) has been developed for rapid free-breathing dynamic MRI (22), and it has been applied to a number of clinical applications (18,23–27). While GRASP ensures good image quality in some body applications such as prostate and breast imaging, it still results in compromised image quality with residual motion blurring for applications with large motion displacement, such as liver and lung imaging (13,28). To address this challenge, GRASP has been extended to XD-GRASP (eXtral-Dimensional GRASP), which performs data sorting based on underlying motion information and then sparse reconstruction to generate a motion-resolved image series to reduce motion blurring (4). XD-GRASP has been applied to both non-DCE-MRI (29–31) and DCE-MRI applications (32), and it has also shown great potential for use in image-guided treatment planning (33). However, like other motion compensation techniques, a major challenge in XD-GRASP is the requirement for reliable motion detection, which could be challenging in some applications such as DCE-MRI. Meanwhile, the need to reconstruct a new respiratory dimension also limits the achievable temporal resolution in DCE-MRI with XD-GRASP.

If temporal resolution is high enough to resolve respiratory motion, 4D real-time MRI (3D+time without any motion-based data sorting) can be a simple way to ultimately address these challenges. For example, given that a standard respiratory cycle typically spans for 4-6 seconds, it is fair enough to assume minimum respiratory motion blurring within each image if one can acquire a 3D image in an order of millisecond. In the same way, 4D real-time imaging with sub-second temporal resolution can capture respiration-induced signal variation and ensure minimized intra-frame blurring. This can also eliminate the need for motion detection that is required in other approaches. Unfortunately, standard MRI, even with the state-of-the-art accelerated imaging techniques available in the clinic, is still not fast enough to achieve this goal. Currently, most real-time MRI applications are performed based on 2D acquisition (34–39), and compromised spatiotemporal resolution is often needed for 3D acquisition (40–43). DCE-MRI also requires real-time acquisition because contrast enhancement is not periodic. However, the achievable temporal resolution in 4D real-time MRI using standard acceleration techniques (e.g., GRASP) is in an order of seconds (23), which is sensitive to intra-frame motion for moving organs such as the liver.

More recently, an improved version of GRASP reconstruction, called GRASP-Pro (44), has been proposed to improve image quality and temporal resolution over GRASP. GRASP-Pro extends GRASP with joint low-rank subspace and sparsity constraints, where the temporal basis used for subspace construction is estimated from an intermediate GRASP reconstruction using low-resolution k-space data. However, with its original implementation, GRASP-Pro still requires sufficient data (radial spokes in each dynamic frame) to ensure good reconstruction performance for the intermediate step, which inherently limits the achievable temporal resolution to fully resolve intra-frame blurring. In this work, we sought to further tailor GRASP-Pro reconstruction for real-time 4D stack-of-stars MRI towards ultra-high temporal resolution below one second. Compared to the original GRASP-Pro implementation and other subspace-based reconstruction methods, the main innovation in our optimization is to estimate the temporal basis from the centers of stack-of-stars k-space. With this simple but effective extension, GRASP-Pro breaks the restrictions associated with the intermediate reconstruction step required in its original implementation; and it is compatible with standard stack-of-stars acquisition. This ultimately leads to a new dynamic MRI framework that enables 4D real-time MRI at sub-second temporal resolution, which intrinsically reduces intra-frame respiratory motion blurring for body applications and eliminates the need for motion detection and motion compensation.

## Methods

### *GRASP-Pro Reconstruction with K-Space Center-Guided Basis Estimation*

With stack-of-stars golden-angle radial sampling, GRASP MRI reconstruction can be implemented in a slice-by-slice matter (either sequential or parallelized), where 2D dynamic image reconstruction is performed for each slice. The use of golden-angle rotation enables flexible data sorting and grouping to reconstruct dynamic images with arbitrary temporal resolution. Standard GRASP implements multicoil compressed sensing reconstruction to solve the following optimization problem:

$$\tilde{\mathbf{m}} = \arg\min_{\tilde{\mathbf{m}}} \frac{1}{2} \| \mathbf{Em} - \sqrt{\mathbf{W}}\mathbf{y} \|_2^2 + \lambda_t \left\| \mathbf{S_t m} \right\|_1 + \lambda_s \left\| \mathbf{S_s m} \right\|_1 \qquad [1]$$

Here, $\mathbf{m} \in \mathbf{C}^{T \times N'^2}$ is the dynamic image series to be reconstructed with an in-plane matrix size of $N \times N$ for each dynamic frame and a number of $T$ dynamic frames. $\mathbf{y}$ is corresponding multicoil dynamic k-space with varying golden-angle radial trajectories sorted for each dynamic frame. $\mathbf{S_t}$ is a temporal sparsifying transform (e.g., temporal finite differences) applied along the dynamic dimension and $\mathbf{S_s}$ is a spatial sparsifying transform (e.g., spatial finite differences) applied to each dynamic frame separately. $\lambda_t$ and $\lambda_s$ are two parameters for the two regularization terms, respectively. In the latest implementation of GRASP reconstruction, self-calibrating GRAPPA Operator Gridding (GROG) (45) is applied to speed up image reconstruction, where $\mathbf{W}$ is a GROG weighting matrix as described in (46). $\mathbf{E} = \sqrt{\mathbf{W}}\mathbf{FC}$ represents the encoding operator for multicoil reconstruction combining coil sensitivities ($\mathbf{C}$), fast Fourier transform (FFT: $\mathbf{F}$) and the GROG weighting function ($\mathbf{W}$).

The use of generic sparisifying transforms to minimize $L_1$ norm in standard GRASP reconstruction limits its achievable acceleration rates, which poses a trade-off between temporal resolution and image quality. To improve reconstruction quality while having high temporal resolution, GRASP has been extended to GRASP-Pro to enforce joint low-rank subspace and sparsity constraints (44). Specifically, GRASP-Pro aims to solve the following optimization problem:

$$\tilde{\mathbf{V}}_{\mathbf{K}} = \arg\min_{\tilde{\mathbf{V}}_{\mathbf{K}}} \frac{1}{2} \| \mathbf{E}(\mathbf{U}_{\mathbf{K}}\mathbf{V}_{\mathbf{K}}) \text{-} \sqrt{\mathbf{W}}\mathbf{y} \|_2^2 + \lambda_1 \left\| \mathbf{S}_{\mathbf{t}}(\mathbf{U}_{\mathbf{K}}\mathbf{V}_{\mathbf{K}}) \right\|_1 + \lambda_2 \left\| \mathbf{S}_{\mathbf{s}}\mathbf{V}_{\mathbf{K}} \right\|_1 \qquad [2]$$

Here, $\mathbf{U}_{\mathbf{K}} \in \mathbf{C}^{T \times K}$ refers to the $K$ dominant basis components taken form the full basis $\mathbf{U} \in \mathbf{C}^{T \times T}$. $\mathbf{V}_{\mathbf{K}} \in \mathbf{C}^{K \times N'^2}$ refers to the spatial basis, or spatial characteristics corresponding to $\mathbf{U}_{\mathbf{K}}$ for representing the dynamic image series to be reconstructed. Optimization of Equation 2 aims to reconstruct $\mathbf{V}_{\mathbf{K}}$, and the dynamic images to be reconstructed ($\mathbf{m}$) are then given as $\mathbf{U}_{\mathbf{K}}\mathbf{V}_{\mathbf{K}} \in \mathbf{C}^{T \times N^2}$. The improved reconstruction performance using GRASP-Pro, compared to standard GRASP, is mainly attributed to the reduced degrees of freedom from the construction of a subspace in iterative reconstruction (47), as one only needs to reconstruct $K$ images ( $\mathbf{V_K}$ ) instead of the original $T$ images ( $\mathbf{m}$ ). An important step to perform subspace-constrained reconstruction is to pre-estimate $\mathbf{U}$ before iterative reconstruction. Although some studies have also proposed to solve both $\mathbf{U_K}$ and $\mathbf{V_K}$ jointly in iterative reconstruction (48,49), this will likely make the optimization problem non-convex and increase the reconstruction complexity substantially.

In the original implementation of GRASP-Pro, $\mathbf{U}$ is pre-estimated in an intermediate reconstruction step using standard GRASP performed on the low-resolution k-space data, as described in (44). However, the need for this intermediate reconstruction step inherently limits achievable acceleration rates and thus achievable temporal resolution, since sufficient k-space data (radial spokes included in each dynamic frame) are still needed for the low-resolution GRASP reconstruction to minimize error in basis estimation. Meanwhile, this extra reconstruction step also requires an additional reconstruction step that prolongs total computation time.

Instead of estimating the temporal basis from an intermediate reconstruction step as described above, we propose to estimate the basis from the k-space centers in stack-of-stars k-space data, which is then used for GRASP-Pro reconstruction. As shown in Figure 1a, stack-of-stars sampling offers a unique feature, with which a number of k-space centers (purple circles) are acquired in each slice location for a radial stack. Here, a radial stack is formed by all the spokes from all slices at a given acquisition angle. While the acquisition angle keeps rotating from one stack to the next, the centers of k-space are repeatedly sampled at the same locations. These k-space centers can be transformed to a series of projections with 1D FFT along the slice dimension to obtain underlying information, such as respiratory motion and/or contrast enhancement, as shown in Figure 1b. These projections can be re-organized to match the target temporal resolution by grouping and averaging consecutive projections together. For example, when every 5 consecutive spokes are combined for image reconstruction, every 5 consecutive projections are averaged for basis estimation. The temporal basis can then be estimated using principal component analysis (PCA):

$$\mathbf{Proj}=\mathbf{UV_{proj}} \qquad [3]$$

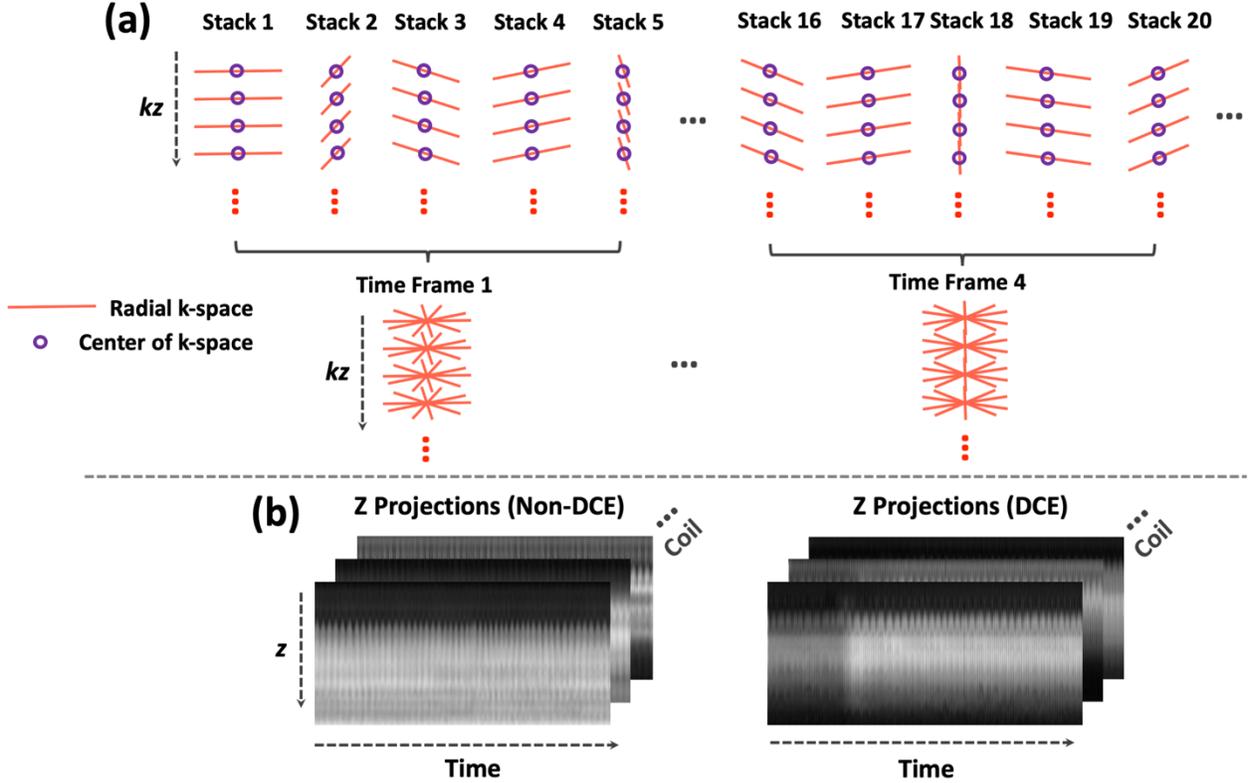

Figure 1: (a) In stack-of-stars radial imaging, the centers of k-space in each radial stack (purple circles) can form a z-directional projection after performing 1D FFT. These projections, when plotted along time, reflect temporal variation for estimating different information, such as respiratory motion and temporal basis functions. In particular, a temporal basis can be used to construct a subspace for low-rank subspace constraint in image reconstruction. When golden-angle radial sampling is implemented, several consecutive spokes can be combined to form a dynamic frame, and a 4D real-time image series can be reconstructed with high temporal resolution. (b) Two examples of z-directional projection profiles, one without contrast injection (left) and one with contrast injection (right).

Here, $\mathbf{Proj} \in \mathbf{C}^{T \times ZC}$ represents concatenated multicoil projections generated from k-space centers of all coil elements, where $Z$ indicates the number of partitions/slices in stack-of-stars k-space data and $\mathbf{V_{proj}} \in \mathbf{C}^{T \times ZC}$ refers to as the coefficients corresponding to the estimated basis. Given the estimated $\mathbf{U}$, the $K$ dominant basis components can be extracted to generate $\mathbf{U_K} \in \mathbf{C}^{T \times K}$ for GRASP-Pro reconstruction as outlined in Equation 2. With this approach, one temporal basis is estimated from the projections for the whole image-set, which can be applied to reconstruct all image slices together or separately.

For simplicity, when GRASP-Pro reconstruction is mentioned hereafter, it is referred to as the optimized reconstruction method with the new basis estimation scheme.

The temporal basis estimated from k-space center projections is used to represent the dynamic image series to be reconstructed. This is valid because: (1) temporal basis contains only temporal information and not spatial information; (2) although the k-space center projections contains limited spatial information, it properly reflects temporal information that is needed for estimating the temporal basis, and (3) the subspace spanned by the temporal basis is not limited to signal evolution in the data used for estimating the basis (k-space center projections); instead, the basis can be used to represent much more signal patterns as long as they are correlated with the training signals (50). Figure 1b shows two representative examples of projections, which provide good information reflecting underlying dynamic changes. It should be noted that the feasibility of estimating temporal basis from 1D projections for subspace-constrained reconstruction has been demonstrated in prior studies (50,51). However, these studies intended to acquire additional 1D navigation data periodically for this purpose, which prolongs overall scan time and needs specific sequence modification. In contrast, the unique sampling feature in the stack-of-stars trajectory enables self-calibration for temporal basis estimation directly from the acquired k-space centers.

### Imaging Experiments

Multiple experiments were designed to test the proposed 4D real-time MRI framework, including numerical motion phantom, free-breathing non-DCE-MRI and DCE-MRI studies in a total of 8 subjects. Lastly, a T1 mapping phantom study was conducted to quantitatively verify the accuracy of basis estimation in GRASP-Pro. Human imaging was HIPAA-compliant and was approved by the local Institutional Review Board (IRB). Written informed consent was obtained from all the subjects prior to their MR scans. MR imaging was performed on 3T clinical MRI scanners (Siemens Magnetom Skyra or Prisma, Erlangen, Germany).

### Abdominal Motion Phantom Study

A realistic 4D abdominal motion phantom was simulated to test highly-accelerated 4D MRI using stack-of-stars golden-angle radial sampling and k-space center-guided GRASP-Pro reconstruction. 4D abdominal phantom images were simulated using source code provided by Lo W et al (52). Imaging parameters for the phantom included: FOV = 380x380mm$^2$, matrix size = 256x256, number of slices = 40, and slice thickness = 5mm. A total of 196 dynamic frames were simulated to cover 7 full respiratory cycles. Assuming a respiratory cycle of ~6 seconds, this led to a temporal resolution of ~210ms/3D volume. To mimic normal breathing in human, the duration of the expiratory phase is longer than the inspiration phase in each respiratory cycle. Ground truth images were first generated after adding zero-mean Gaussian noise with a signal-to-noise ratio (SNR) of 46dB to the simulated images. Golden-angle stack-of-stars k-space data were then synthesized for each image slice, where each dynamic frame contains 5 consecutive spokes. The rotation angle is set the same for each radial stack to simulate standard stack-of-stars sampling. For each slice, dynamic images were reconstructed using non-uniform FFT (NUFFT), standard GRASP, and GRASP-Pro. For additional comparison, GRASP-Pro reconstruction was also performed using a temporal basis estimated from an intermediate GRASP reconstruction (the original implementation, denoted as GRASP-Pro-orig) and from the ground truth images (denoted as GRASP-Pro-ref), respectively. It should be noted that GRASP-Pro-ref is only available in simulated dataset, since fully sampled references are not available in practice. Root mean square error (RMSE) and structural similarity index (SSIM) were calculated for each dynamic frame with respect to the ground truth.

### *Non-DCE-MRI Study*

4D real-time MRI was tested in 5 stack-of-stars golden-angle radial abdominal MRI datasets (referred to as non-DCE dataset 1-5) acquired in 4 subjects without contrast injection.

Non-DCE dataset 1 was acquired in subject 1 (male, age=28) in the axial orientation. Relevant imaging parameters included: Matrix size = 256x256, FOV = 350x350mm$^2$, in-plane resolution = 1.37x1.37mm$^2$, slice thickness = 5mm, TR/TE = 3.62/1.57ms, flip angle (FA) = 12°, number of spokes acquired in each slice = 650. With 80% slice partial Fourier (PF), 38 slices were acquired for each radial stack. The total scan time was 102 seconds. For the first comparison, both GRASP reconstruction and GRASP-Pro reconstruction were performed on the dataset to generate 4D real-time images with 5 consecutive spokes grouped in each dynamic frame in each slice. This resulted in a temporal resolution of 0.78 seconds/3D volume (102/650*5). For the second comparison, XD-GRASP was performed on the dataset to reconstruct 4 respiratory motion-resolved images spanning from end-expiration to end-inspiration, as described previously (4). The number of spokes for each respiratory phase in the XD-GRASP reconstruction was 162 (650/4), and thus, the acceleration rate was ~32-fold lower compared to GRASP-Pro.

Non-DCE dataset 2 was acquired in subject 2 (female, age=32) in the axial orientation. Relevant imaging parameters included: Matrix size = 192x192, FOV = 300x300mm$^2$, in-plane resolution = 1.56x1.56mm$^2$, slice thickness = 5mm, TR/TE = 3.52/1.49ms, FA = 12$^o$, number of spokes acquired in each slice = 800. With 80% slice PF, the number of acquired slices was 28. The total scan time was 100 seconds. Non-DCE dataset 3 was acquired in subject 3 (male, age=24) in the coronal orientation. Relevant imaging parameters included: Matrix size = 192x192, FOV = 340x340mm$^2$, in-plane resolution = 1.77x1.77mm$^2$, slice thickness = 5mm, TR/TE = 4.18/1.49ms, number of spokes acquired in each slice = 1200. With 80% slice PF, the number of acquired slices was 26. The total scan time was 163 seconds. GRASP-Pro reconstruction was performed on these two datasets to generate 4D real-time images with 5 consecutive spokes grouped in each dynamic frame in each slice. This resulted in a temporal resolution of 0.63 seconds/3D volume (100/800*5) for dataset 2 and 0.68 seconds/3D volume (163/1200*5) for dataset 3.

Non-DCE dataset 4 and 5 were both acquired in subject 4 (male, age=27), one in the axial orientation (dataset 4) and the other one in the coronal orientation (dataset 5). For acquisition of both datasets, the volunteer was asked to perform consistent deep breathing as much as possible. Relevant imaging parameters were: Matrix size = 192x192, FOV = 360x360mm$^2$, in-plane resolution = 1.88x1.88mm$^2$, slice thickness = 5mm, TR/TE = 3.43/1.43ms, FA = 12$^o$, number of spokes acquired in each slice = 800. With 80% slice PF, the number of acquired slices was 20. The total scan time was 70 seconds. Relevant imaging parameters for dataset 5 included: Matrix size = 192x192, FOV = 360x360mm$^2$, in-plane resolution = 1.88x1.88mm$^2$, slice thickness = 5mm, TR/TE = 3.43/1.43ms, FA = 12$^o$, number of spokes acquired in each slice = 1304. With 80% slice PF, the number of acquired slices was 28. The total scan time was 159 seconds. GRASP-Pro reconstruction was performed on these two datasets to generate 4D real-time images with 5 consecutive spokes grouped in each dynamic frame in each slice. This resulted in a temporal resolution of 0.44 seconds/3D volume (70/800*5) and 0.61 seconds/3D volume (159/1304*5).

*DCE-MRI Study*

4D real-time MRI was also tested in 4 stack-of-stars golden-angle radial MRI datasets acquired in 4 subjects following the injection of a Gadolinium-based contrast agent (referred to as DCE dataset 1-4).

DCE dataset 1 was acquired in the liver of subject 1 (female, age=28) in the axial orientation. Relevant imaging parameters included: Matrix size = 256x256, FOV = 350x350mm$^2$, in-plane resolution = 1.37x1.37mm$^2$, slice thickness = 5mm, TR/TE = 3.6/1.6ms, FA = 12$^o$, number of spokes acquired in each slice = 1222. With 80% slice partial Fourier, 38 splices were acquired for each radial stack. The total scan time was 190 seconds. For the first comparison, GRASP and motion-weighted GRASP reconstruction (13) was first performed with 96 consecutive spokes grouped in each dynamic frame in each slice with a temporal resolution = 15 seconds/3D volume. GRASP-Pro reconstruction was then performed with 5 consecutive spokes grouped in each dynamic frame in each slice with a temporal resolution of 0.78 seconds/3D volume (190/1222*5). For the second comparison, the temporal resolution was further pushed to 0.47 seconds/3D volume with only 3 consecutive spokes grouped in each dynamic frame in each slice. The GRASP-Pro reconstruction was additionally compared with the low-rank plus sparsity (L+S) reconstruction (53) using MATLAB code provided by the authors.

DCE dataset 2 was acquired in the lung of subject 2 (female, age=34) in the axial orientation. Relevant imaging parameters included: Matrix size = 256x256, FOV = 350x350mm$^2$, in-plane resolution = 1.37x1.37mm$^2$, slice thickness = 5mm, TR/TE = 3.5/1.55ms, FA = 12$^o$, number of spokes acquired in each slice = 1200. With 80% slice partial Fourier, 38 splices were acquired for each radial stack. The total scan time was 182 seconds. GRASP-Pro reconstruction was performed on the dataset to generate 4D real-time images with 5 consecutive spokes grouped in each dynamic frame in each slice with a temporal resolution of 0.76 seconds/3D volume (182/1200*5).

DCE dataset 3 was acquired in the prostate of subject 3 (male, age=61) in the axial orientation. Relevant imaging parameters included: Matrix size = 224x224, FOV = 240x240mm$^2$, in-plane resolution = 1.07x1.07mm$^2$, slice thickness = 5mm, TR/TE = 4.12/1.96ms, FA = 12$^o$, number of spokes acquired in each slice = 1755. With 80% partial Fourier applied along the slice dimension, 24 splices were acquired for each radial stack. The total scan time was 221 seconds. GRASP-Pro reconstruction was performed on the dataset to generate 4D real-time images with 5 consecutive spokes grouped in each dynamic frame in each slice with a temporal resolution of 0.63 seconds/3D volume (221/1755*5).

DCE dataset 4 was performed in the liver of subject 4 (female, age=57) in the axial orientation. Relevant imaging parameters were the same as that for acquiring DCE dataset 1. The subject had irregular breathing pattern during the MR imaging. GRASP-Pro reconstruction was performed on the dataset to generate 4D real-time images with 5 consecutive spokes combined in each dynamic frame in each slice with a temporal resolution of 0.78 seconds/3D volume (190/1222*5). For additional comparison, GRASP reconstruction and motion-weighted GRASP reconstruction were also performed in this dataset both with a temporal resolution of 15 seconds/3D volume (96 consecutive spokes in each dynamic frame in each slice).

To assess the temporal fidelity of compressed sensing reconstruction in DCE-MRI, we previously proposed to use NUFFT as an indirect reference (22). This is based on the observation that without any temporal regularization, NUFFT images can still preserve the true temporal fidelity despite the presence of streaking artifacts, especially with region-of-interest (ROI)-based analysis. In this study, NUFFT images were also used as a reference to assess the temporal fidelity of GRASP-Pro reconstruction. However, given that NUFFT reconstruction with high acceleration rates has substantial streaking artifacts, it was performed with a lower temporal resolution compared to GRASP-Pro to reduce the influence of streaking artifacts.

*T1 Mapping Phantom Study*

To quantitatively verify the accuracy of GRASP-Pro reconstruction using k-space center-guided temporal basis estimation, a T1 mapping phantom dataset was acquired using inversion-recovery prepared stack-of-stars sampling as described in (54). The dataset was reconstructed using GRASP-Pro reconstruction with temporal basis estimated from the Bloch simulation (served as reference) and k-space centers, respectively.

*Implementation of Image Reconstruction*

All image reconstruction tasks were performed in MATLAB (MathWorks, MA) with GPU implementation by using the "gpuArray" function. Radial k-space data were prepared using the unstreaking technique in a pre-processing step to improve image quality (13,55). Self-calibrating GROG was implemented to replace NUFFT for iterative reconstruction to improve reconstruction speed (45,46).

## Results

*Abdominal Motion Phantom Study*

Results comparing NUFFT, GRASP, GRASP-Pro reconstruction with different basis estimation schemes and the ground truth for one slice of the abdominal motion phantom, along with corresponding x-t temporal profiles, are presented in Figure 2. With only 5 spokes per dynamic frame, NUFFT reconstruction yielded substantial streaking artifacts and GRASP failed to remove those artifacts. GRASP-Pro reconstruction, in comparison, was able to generate substantially improved image quality and there is no visual difference between GRASP-Pro and GRASP-Pro-ref. While GRASP-Pro-orig can still achieve much improved image quality compared to GRASP, the result suffers from residual streaking artifacts. This may be because the basis for subspace construction in GRASP-Pro-orig was estimated from an intermediate reconstruction step on low-resolution k-space using GRASP, and the performance of GRASP can be limited at this accelerated rate. Thus, errors generated in basis estimation are likely propagated to the final reconstructed images. The RMSE and SSIM results indicate that the expiratory motion phases can reconstructed better than the inspiratory phases. This is likely due to reduced temporal correlations in inspiratory phase compared to expiratory phase as simulated in the phantom dataset.

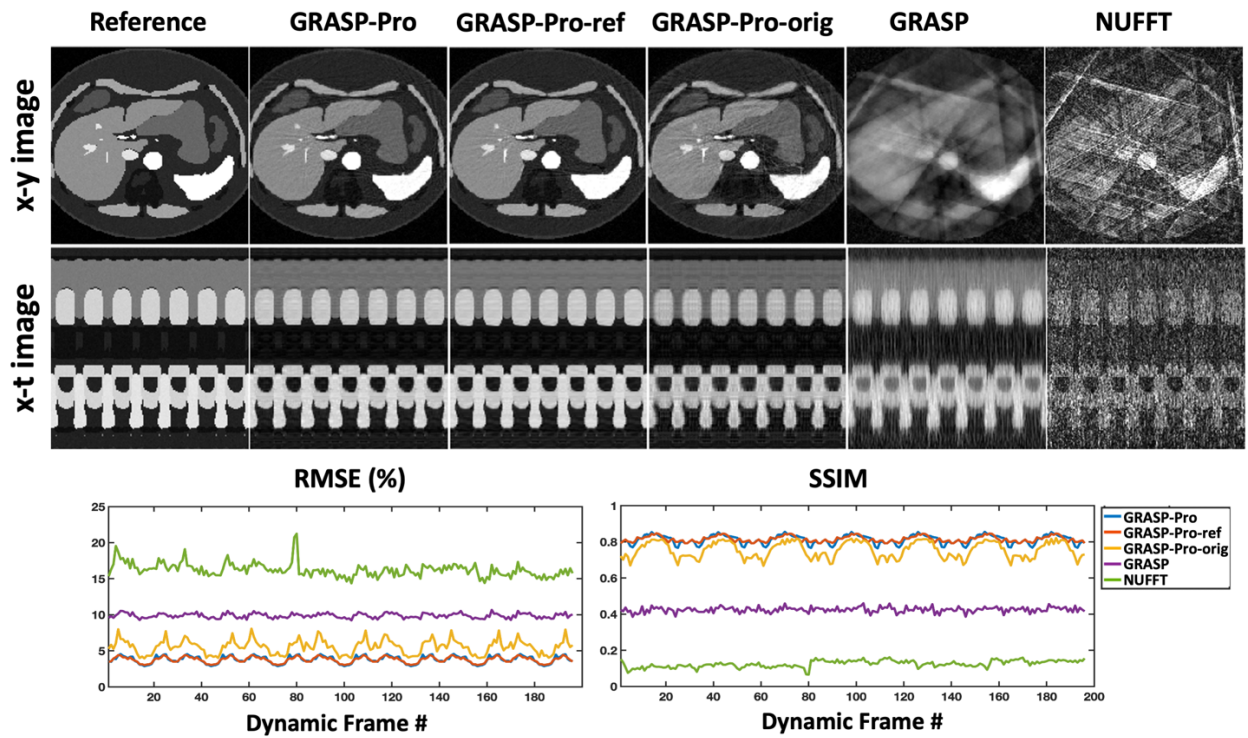

**Figure 2.** (a) Comparison of NUFFT, GRASP and GRASP-Pro reconstruction with different basis estimation schemes (GRASP-Pro-ref: from fully sampled images; GRASP-Pro: from stack-of-stars k-space centers; GRASP-Pro-orig: from an intermediate reconstruction step). Image reconstruction was performed with 5 spokes grouped in each dynamic frame in each slice using the stack-of-stars golden-angle radial trajectory. NUFFT reconstruction yielded substantial undersampling artifacts. GRASP reconstruction failed to remove artifacts. GRASP-Pro-ref and GRASP-Pro reconstruction were able to recover images with comparable image quality with respect to the reference. GRASP-Pro-orig achieved compromised reconstruction performance. The RMSE and SSIM for each dynamic frame were plotted for each reconstruction method.

### *Non-DCE-MRI Study*

Results comparing NUFFT, GRASP and GRASP-Pro reconstruction for 4D real-time MRI with a temporal resolution of 0.78 seconds/3D volume are shown for non-DCE dataset 1 in Figure 3. NUFFT reconstruction produced strong streaking artifacts at this accelerate rate. Although GRASP could remove many of those artifacts, the reconstruction quality remains low. In comparison, GRASP-Pro was able to generate clean and sharp images with good delineation of image structures in different dynamic frames.

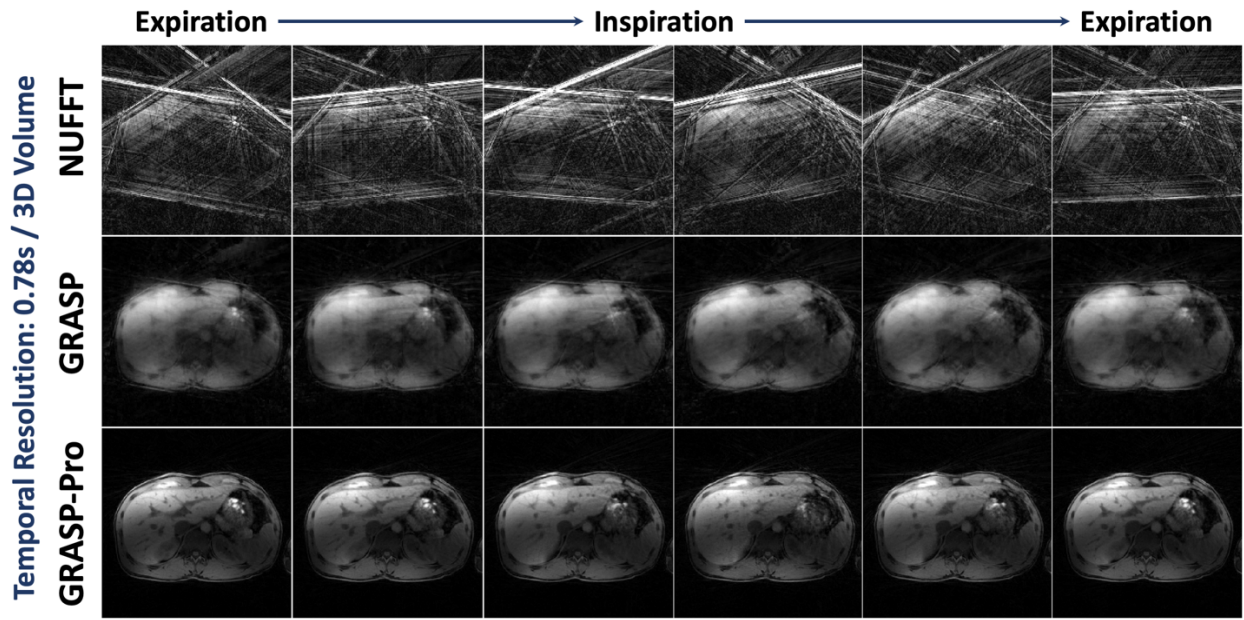

**Figure 3.** Comparison of NUFFT, GRASP and GRASP-Pro reconstruction for 4D real-time MRI (one representative slice) for non-DCE dataset 1. For each slice of the stack-of-stars dataset, image reconstruction was performed with only 5 spokes in each temporal frame, leading to a temporal resolution of 0.78 seconds/3D volume. NUFFT reconstruction produced substantial streaking artifacts. Although GRASP could remove some artifacts, the reconstruction quality remains low. In comparison, GRASP-Pro was able to reconstruct clean and sharp images with good delineation of image structures in different temporal frames.

For the same dataset, Figure 4 compares motion-averaged images reconstructed using NUFFT (all 650 spokes), GRASP-Pro with 5 spokes grouped in each dynamic frame (a total of 130 dynamic frames) and XD-GRASP with 4 respiratory-resolved phases (162 spokes in each phase). Compared to the motion-averaged NUFFT reconstruction, both GRASP-Pro and XD-GRASP reconstruction yielded improved image sharpness as indicated by the better delineation of vessel-tissue boundaries. However, the inspiratory images reconstructed with XD-GRASP is slightly better than corresponding images reconstructed with GRASP-Pro, while the expiratory images are more comparable. This is consistent with that observed in the motion phantom study (see Figure 2) considering the fact that the expiratory duration is normally longer in human breathing pattern.

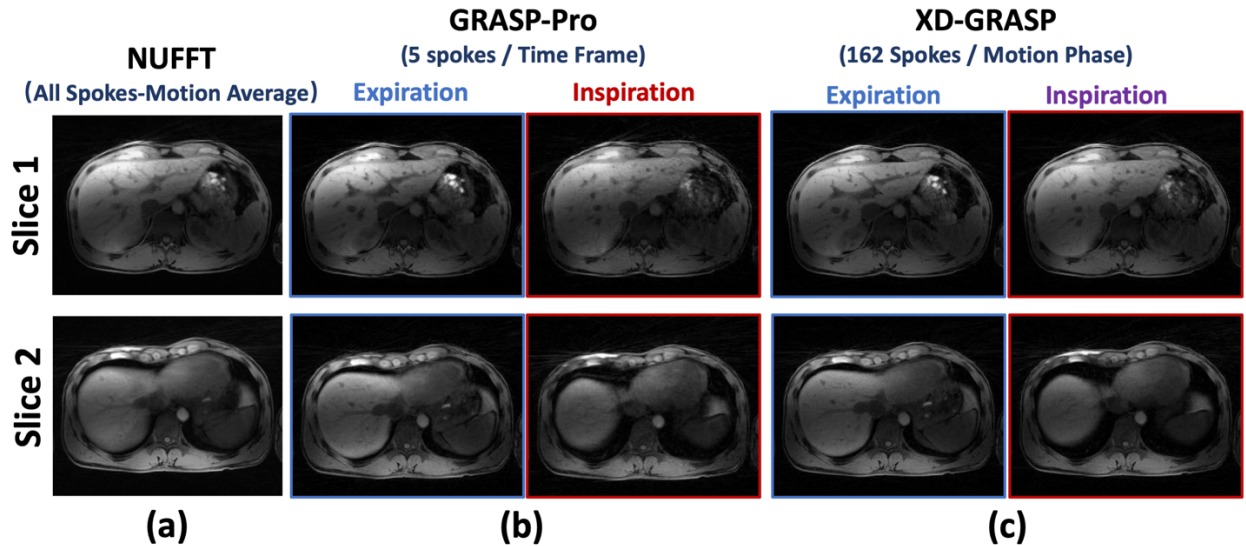

**Figure 4.** Comparison of motion-averaged NUFFT reconstruction using all the acquired spokes, 4D real-time images reconstructed using GRASP-Pro (5 spokes each temporal frame) and 4D respiratory-resolved images reconstructed using XD-GRASP (162 spokes each respiratory phase) in non-DCE dataset 1. Both expiratory and inspiratory images are displayed for GRASP-Pro and XD-GRASP reconstruction. Compared to motion-averaged reconstruction, both GRASP-Pro and XD-GRASP reconstruction yielded improved image sharpness as indicated by the better delineated vessel-tissue boundaries.

Figure 5 shows 4D real-time MR images reconstructed using GRASP-Pro in non-DCE dataset 2 (axial imaging) and dataset 3 (coronal imaging). Images are displayed for an expiratory phase and an inspiratory phase from 2 slices. With the sub-second temporal resolution (0.63 seconds/3D volume and 0.68 seconds/3D volume), intra-frame blurring is reduced without the need of explicit motion compensation, while good image quality can be preserved. Figure 6 shows 4D real-time MR images reconstructed using GRASP-Pro in non-DCE dataset 4 and dataset 5 with consistent deep breathing. Images are displayed for an expiratory phase and an inspiratory phase. With a temporal resolution of 0.44 seconds/3D volume and 0.61 seconds/3D volume, strong motion blurring is presented in the motion-averaged images, which can be largely reduced with GRASP-Pro reconstruction. However, it was noted that the overall image quality of these two datasets is degraded compared to normal breathing.

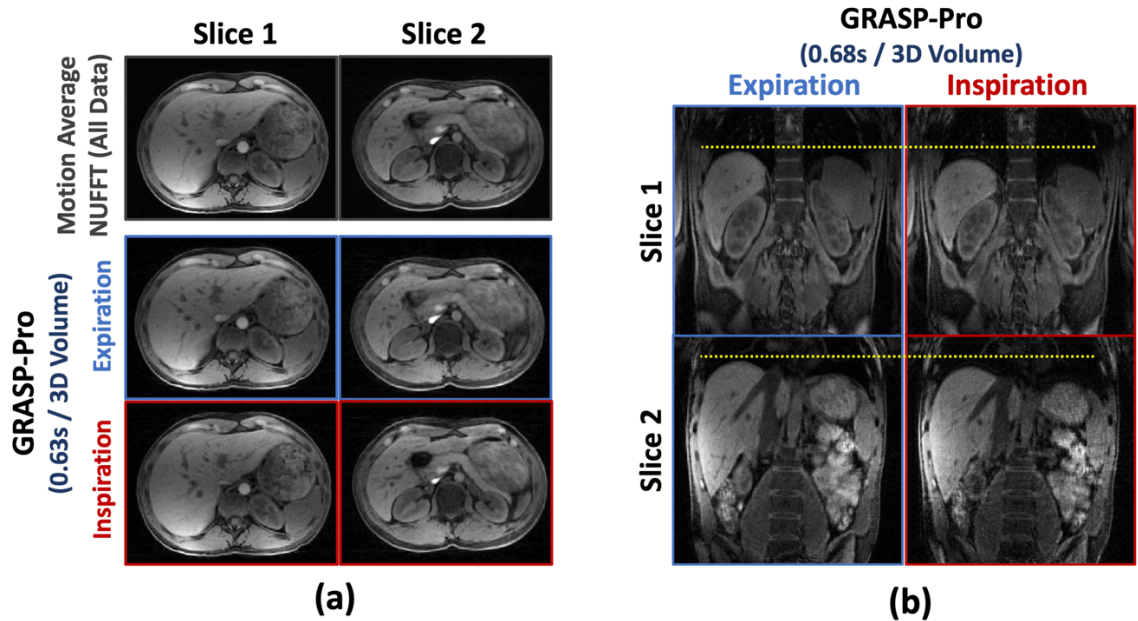

**Figure 5**. 4D real-time images (both expiratory and inspiratory phases) reconstructed using GRASP-Pro for non-DCE dataset 2 and non-DCE dataset 3 with 5 spokes grouped in each temporal frame in each slice. The temporal resolution is 0.63 seconds/3D volume (axial imaging) and 0.68 seconds/3D volume (coronal imaging).

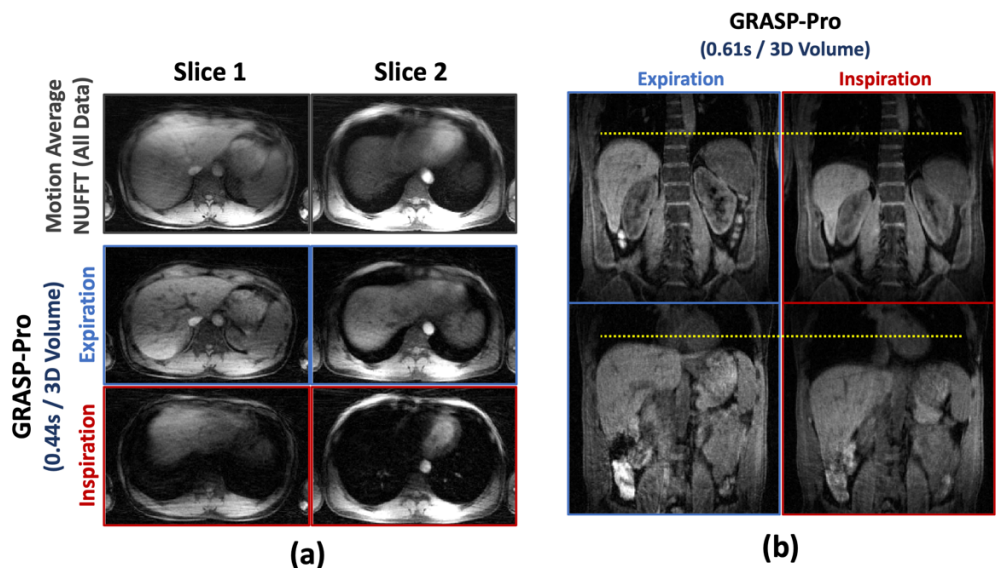

**Figure 6**. GRASP-Pro reconstruction in non-DCE dataset 4 and non-DCE dataset 5 with consistent deep breathing. The temporal resolution is 0.44 seconds/3D volume for the axial images and is 0.61 seconds/3D volume for the coronal images. Although the overall image quality is reduced compared to normal breathing (non-DCE dataset 1-3), GRASP-Pro was still able to reconstruct good images, especially compared to the motion-averaged images with strong respiratory blurring.

*DCE-MRI Study*

Figure 7 compares standard GRASP reconstruction and motion-weighted GRASP reconstruction with a temporal resolution of 15 seconds/3D volume with GRASP-Pro reconstruction with sub-second temporal resolution (0.78 seconds/3D volume) in DCE dataset 1. GRASP-Pro reconstruction did not implement any motion compensation algorithm and the expiratory phases in different contrast enhanced phases are displayed. Despite much more aggressive acceleration, GRASP-Pro reconstruction yielded good reconstruction quality that is comparable to motion-weighted GRASP reconstruction. Standard GRASP reconstruction suffers from residual motion blurring as indicated by the blurred hepatic vessels (yellow arrows). More contrast phases from different slices reconstructed with GRASP-Pro in this dataset are also shown in Figure 8.

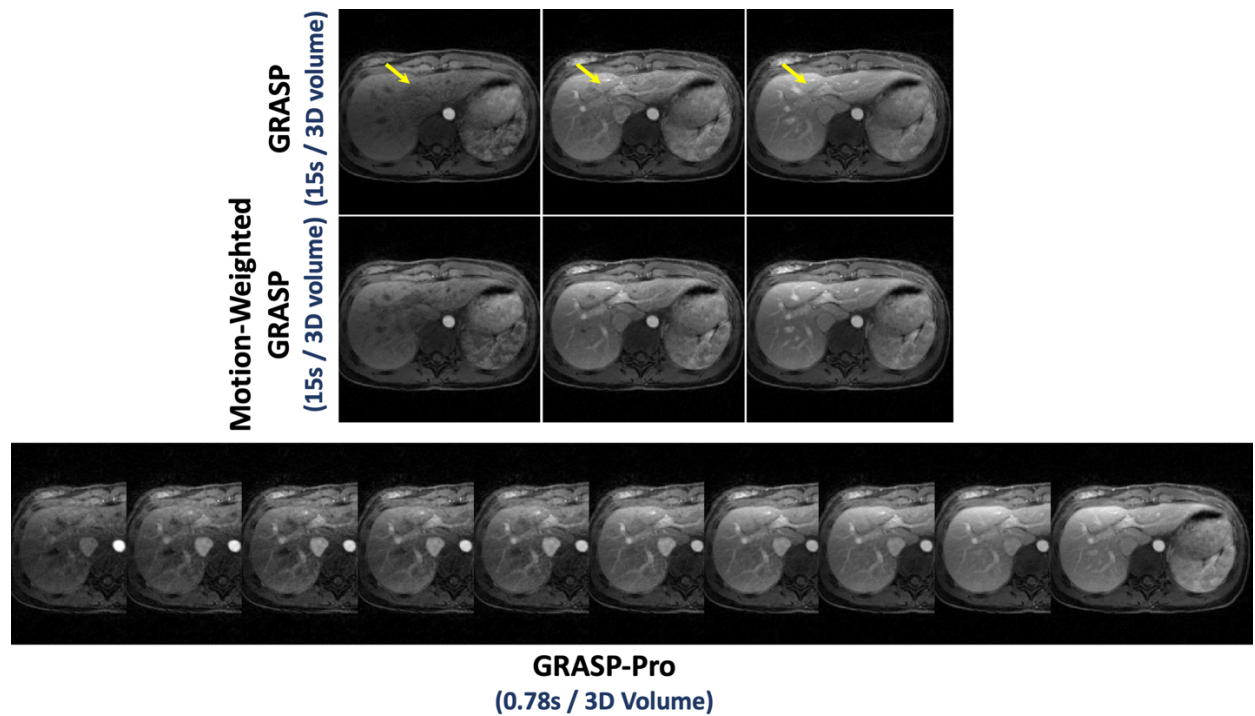

**Figure 7**. Comparison of low-temporal-resolution GRASP reconstruction (15 seconds/3D volume) and motion-weighted GRASP reconstruction (15 seconds/3D volume) with high-temporal-resolution GRASP-Pro reconstruction (0.78 seconds/3D volume) in DCE dataset 1. GRASP-Pro reconstruction did not implement any motion compensation algorithm and only expiratory phases are displayed. Despite substantially higher temporal resolution (0.78 seconds/3D volume verse 15 seconds/3D volume), GRASP-Pro reconstruction still led to good reconstruction quality compared to motion-weighted GRASP

reconstruction, and standard GRASP reconstruction suffers from residual motion blurring as indicated by the blurred hepatic vessels (yellow arrows).

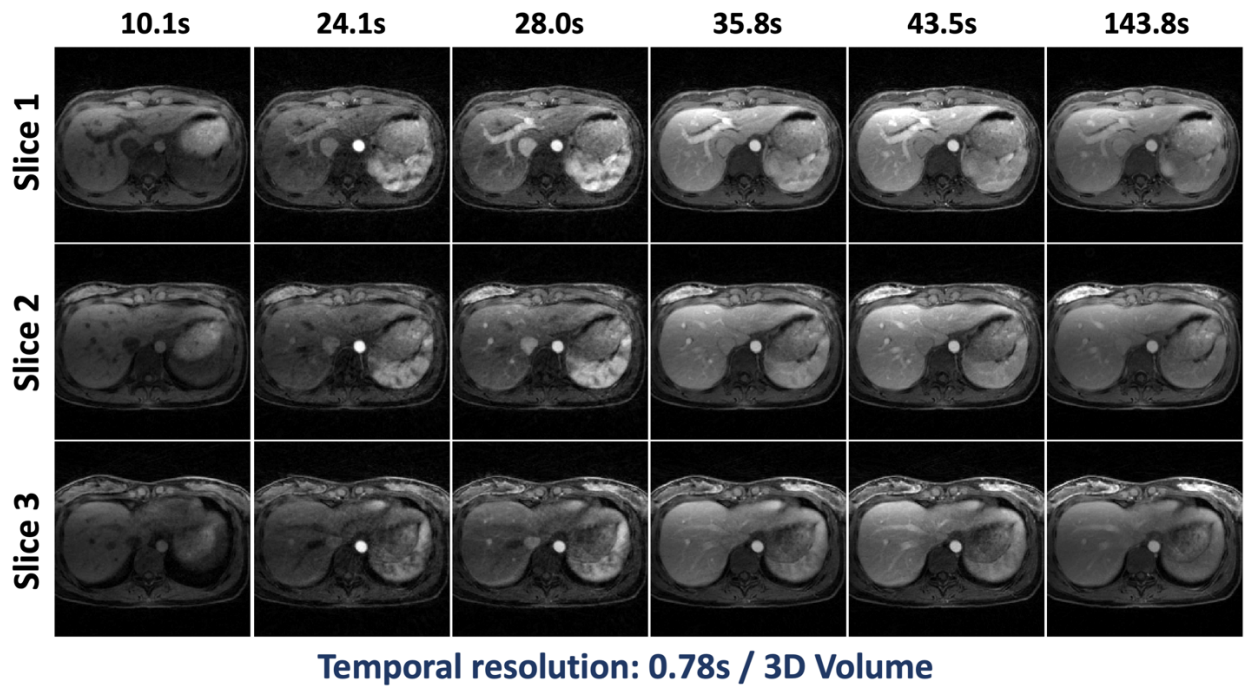

**Temporal resolution: 0.78s / 3D Volume**

**Figure 8**. Three slices of different contrast phases in DCE dataset 1 reconstructed using GRASP-Pro with a temporal resolution of 0.78 seconds/3D volume.

Figure 9 compares NUFFT, GRASP, L+S and GRASP-Pro reconstruction for 4D real-time MRI with a temporal resolution of 0.47 seconds/3D volume in DCE dataset 1. Images are displayed for a pre-contrast phase, an arterial phase, a venous phase and a delayed phase. Compared to different reference method, GRASP-Pro was able to reconstruct clean and sharp images with good delineation of image structures in different contrast-enhanced phases despite vast acceleration.

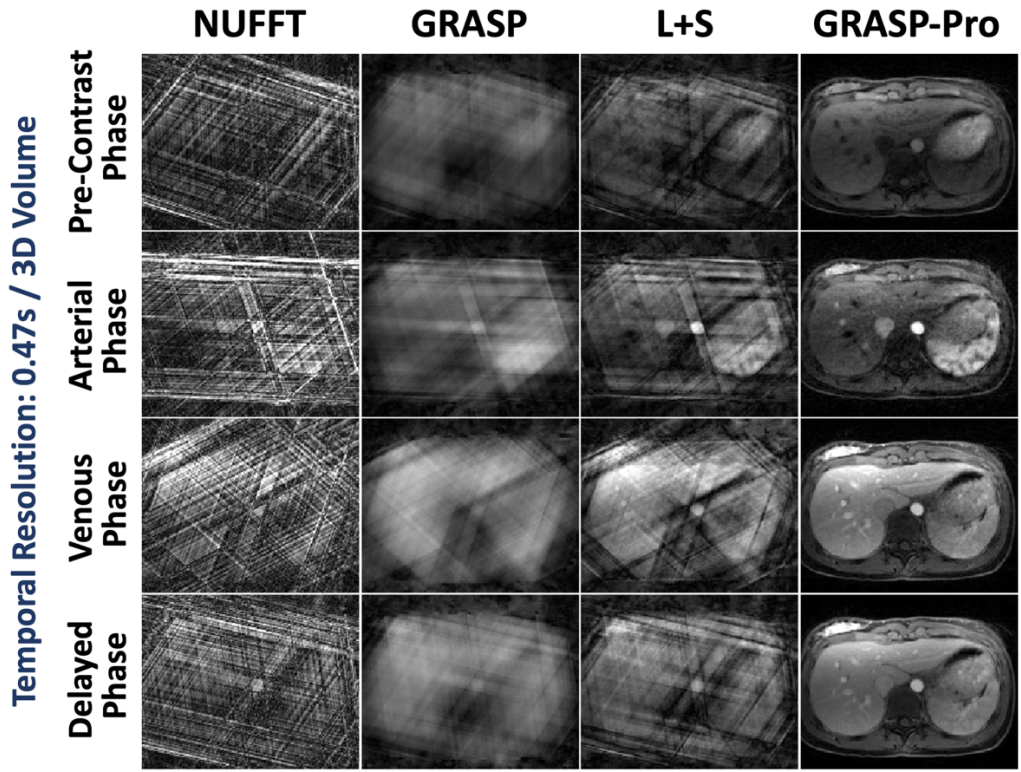

**Figure 9**. Comparison of NUFFT, GRASP, L+S and GRASP-Pro reconstruction for real-time 4D MRI (one representative slice) for DCE dataset 1. For each slice of the stack-of-stars dataset, image reconstruction was performed with only 3 spokes in each temporal frame, leading to a temporal resolution of 0.47 seconds/3D volume. NUFFT reconstruction produced strong streaking artifacts. Although GRASP and L+S could remove some artifacts, the reconstruction quality remains low. In comparison, GRASP-Pro was able to reconstruct clean and sharp images with good delineation of image structures in different contrast-enhanced phases.

Figure 10 shows GRASP-Pro reconstruction in DCE dataset 2 (upper panel, lung imaging) and DCE dataset 3 (lower panel, prostate imaging). The results clearly demonstrate the recovery of image quality from NUFFT reconstruction with 5 spokes per temporal frame in each slice. Figure 11 compares the contrast enhancement curves in different ROIs between NUFFT and GRASP-Pro reconstruction in DCE dataset 1-3 (a-c). As described in the method section, the temporal resolution of the NUFFT reconstruction was set much lower than GRASP-Pro to have sufficient image quality as a reference for assessing temporal fidelity. That is why the NUFFT curves show larger step size and some temporal blurring.

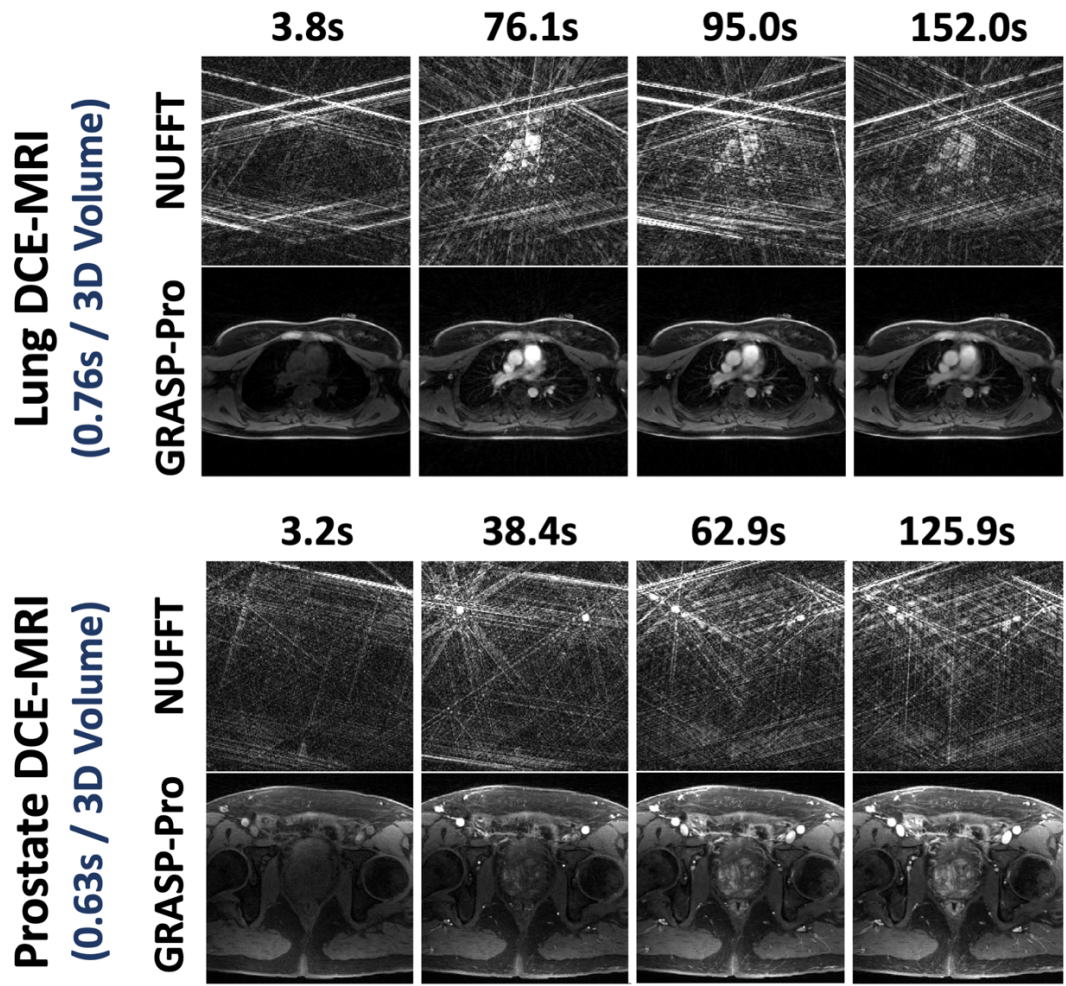

**Figure 10**. GRASP-Pro reconstruction in DCE dataset 2 (upper panel, lung imaging) and DCE dataset 3 (lower panel, prostate imaging) showing the recovery of image quality from NUFFT reconstruction with 5 spokes per temporal frame in each slice. The temporal resolution is 0.76 seconds/3D volume for the lung images and is 0.63 seconds/3D volume for the prostate images.

Figure 12 shows images reconstructed with GRASP-Pro in DCE dataset 4. While the subject had irregular breathing during data acquisition (as shown by the projections in Figure 10a), GRASP-Pro was able to reconstructed good images in different contrast phases with a temporal resolution of 0.78 seconds/3D volume in each slice, which clearly differentiate different motion phases in a respiratory cycle (Figure 10b). Comparison of GRASP and motion-weighted GRASP reconstruction with a temporal resolution of 15 seconds/3D volume with GRASP-Pro reconstruction with sub-second temporal resolution (0.78 seconds/3D volume) in a venous phase from this dataset is shown in Figure 10c.

The GRASP image shows residual blurring. While motion-weighted reconstruction improved the overall image quality, the delineation of small vessels is still slightly compromised compared to GRASP-Pro, as indicated by the yellow arrows, which is likely due to residual intra-frame motion blurring.

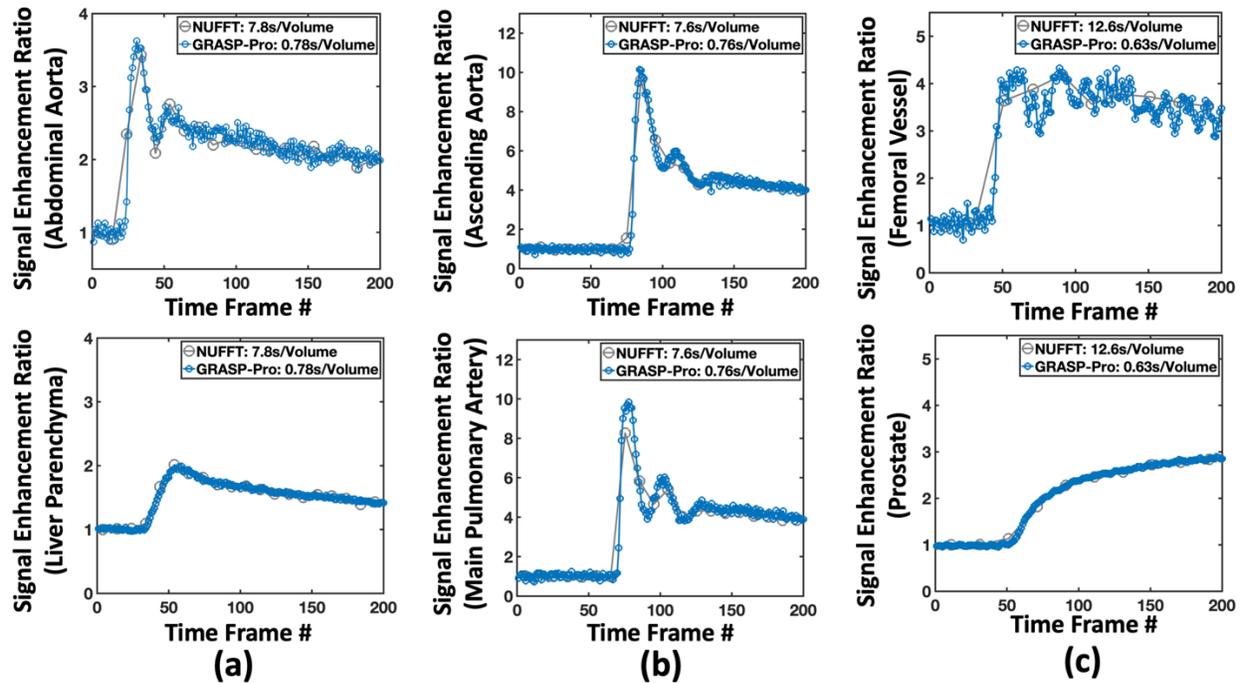

**Figure 11.** Contrast enhancement curves from the 4D DCE-MRI of the liver, the lung and the prostate reconstructed using GRASP-Pro DCE datasets 1-3. For comparison and validation, corresponding contrast enhancement curves from the 4D DCE-MRI reconstructed using NUFFT with a lower temporal resolution (as indicated in the figure legend) are also displayed. Although the NUFFT images have streaking artifacts, temporal fidelity is expected preserved in ROI-averaged contrast enhancement curves.

### *T1 Mapping Phantom Study*

Finally, Figure 13 compares T1 maps generated from images reconstructed with GRASP-Pro using a temporal basis estimated from a dictionary generated using the Bloch equations and a temporal basis estimated from the centers of k-space. There is no visual difference in the T1 maps, which is confirmed by the quantitative analysis of mean T1 values in different phantom vials. The linear correlation analysis yielded an $R^2$ of 0.999, and the Lin's Concordance Correlation Coefficient (CCC) was excellent (0.9994). These results have quantitatively shown that k-space center-guided basis estimation can be accurate for subspace reconstruction.

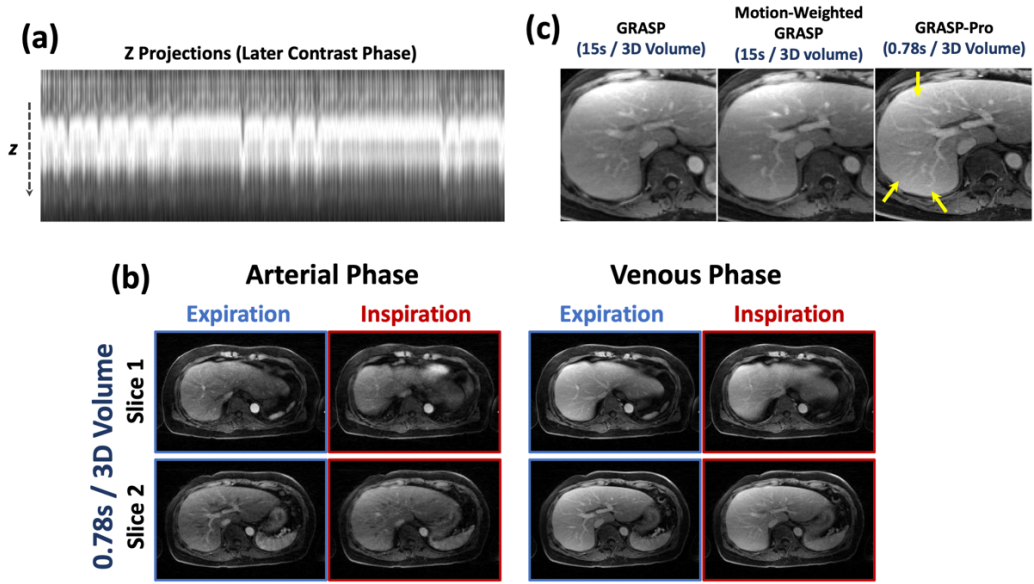

**Figure 12.** GRASP-Pro reconstruction in DCE dataset 4. (a) the z projection profiles show that the subject had irregular breathing during data acquisition. (b) GRASP-Pro was able to reconstructed good images in different contrast phases with a temporal resolution of 0.78 seconds/3D volume. (c) Comparison of low-temporal-resolution GRASP reconstruction (15 seconds/3D volume) and motion-weighted GRASP reconstruction (15 seconds/3D volume) with high-temporal-resolution GRASP-Pro reconstruction (0.78 seconds/3D volume) in a venous contrast phase. GRASP-Pro was able to recover small details (small vessels) that are blurred in GRASP and motion-weighted GRASP (likely due to motion and irregular breathing from low-temporal-resolution reconstruction).

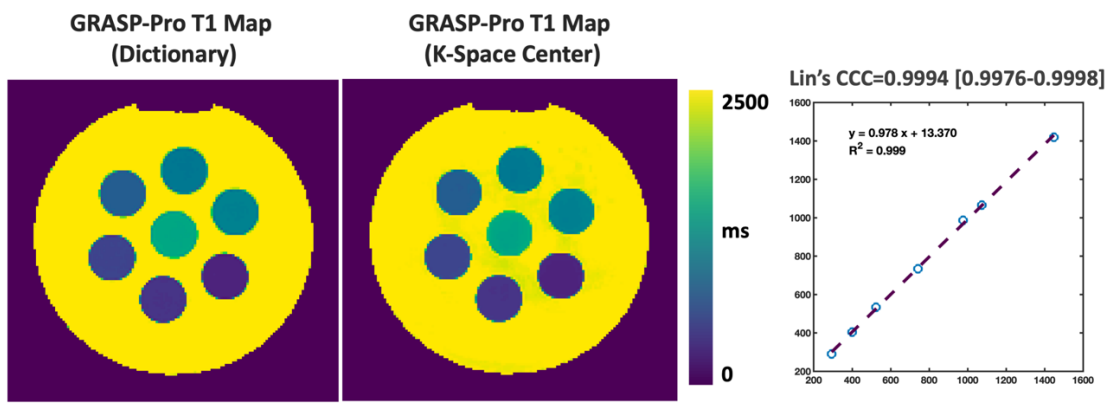

**Figure 13.** Comparison of T1 maps generated from images reconstructed with GRASP-Pro using a temporal basis estimated from a dictionary generated using the Bloch equations and a temporal basis estimated from the centers of k-space. There is no visual difference in the T1 maps, which is confirmed by the quantitative analysis of mean T1 values in different phantom vials.

## Discussion

In this study, a dynamic MRI framework is proposed, which enables 4D real-time imaging at sub-second temporal resolution. This is achieved with a combination of stack-of-stars golden-angle radial sampling and GRASP-Pro reconstruction that is tailored for temporal basis estimation. With GPU implementation, the time need to reconstruct each image slice was about ~2-3 minutes. The performance of our imaging technique has been demonstrated in different experiments in numerical motion phantom and multiple in-vivo datasets, and it has been compared against state-of-the-art motion-compensation methods based on respiratory motion-resolved reconstruction and motion-weighted reconstruction. The significance of performing 4D real-time MRI at sub-second temporal resolution goes beyond pushing imaging speed. Importantly, it provides a simple and effective way to perform fast free-breathing dynamic imaging that does not require specific sequence modification, does not require motion detection, and can intrinsically address intra-frame respiratory blurring.

Sparse image reconstruction with subspace constrains can be performed in different ways. The temporal basis that is needed for subspace construction can either be pre-estimated before iterative reconstruction or jointly reconstructed together with the to-be-reconstructed image series. Pre-estimation of the basis has some advantages, as it can simplify and accelerate the reconstruction process compared to joint basis-image reconstruction, which tends to be non-convex, complex, and slow. The pre-estimation of the basis has been proposed based on (a) fully-sampled low-resolution k-space region (47), (b) additionally acquired navigation data (50), or (c) an intermediate reconstruction step on undersampled low-resolution k-space region (44). The estimation scheme (a) has several major limitations, including the need for specific sequence modification and prolonged scan time to acquire k-space center, which limits achievable acceleration rate and temporal resolution. Meanwhile, it is mainly implemented for Cartesian sampling and can become problematic in others such as radial sampling. The main limitation of the estimation scheme (b) is the need of additional time to acquire navigation data for basis estimation, and this requires specific modification to the imaging sequence. The estimation scheme (c) was proposed in the original implementation of GRASP-Pro reconstruction. It is simple and does not need sequence adaption. However, the main restriction of this approach is the need to have sufficient data for the intermediate reconstruction. Any reconstruction errors, such as temporal blurring or residual undersampling artifacts, can be propagated to the final reconstructed images. As a result, this requirement limits achievable temporal resolution as shown in our experiments (Figure 2). To overcome these limitations, the main contribution of this study is to show that one can estimate a temporal basis from the centers of k-space in stack-of-stars imaging. This simple extension nicely breaks the restrictions in estimation scheme (c) and enables highly-accelerated data acquisition and thus exquisite temporal resolution to address intra-frame motion blurring.

Stack-of-stars sampling is ideal to perform the proposed 4D real-time imaging, as the k-space centers are repeatedly sampling to form z-directional projections for basis estimation. Therefore, it provides a nature way for subspace-constrained reconstruction without the need for sequence modification, which facilitates easy clinical translation and dissemination. The use of k-space centers to guide image reconstruction in stack-of-stars imaging has been demonstrated in many prior studies, but the main use has been limited to motion detection, self-gating or motion sorting (4,10). For motion detection, the z-directional projections need to be further processed (e.g., filtering) to generate a 1D motion signal (4,10). Thus, the performance is highly dependent on the underlying breathing pattern and can be less reliable with the underlying contrast changes (e.g., in DCE-MRI). In comparison, estimation of a temporal basis from z-directional projections can be more robust, since the basis is estimated from the entire projections without additional processing.

Our study shares some similarities with two recent works on highly-accelerated DCE-MRI. The first one is called extreme MRI (49), which enables 4D real-time MRI at sub-second temporal resolution using 3D radial sampling and a multiscale low-rank subspace constraint. However, this approach aims to reconstruct basis and images jointly and requires a specific reconstruction algorithm to solve a non-convex optimization problem for 4D MRI reconstruction, which is slow (e.g., 6-42 hours) for clinical translation. Meanwhile, 3D radial trajectories usually require data acquisition with isotropic resolution. Although this is desired in whole-heart imaging, it may not be necessary in many other applications and the computational burden is very high (14). The second one is based on the MR multitasking technique for fast DCE-MRI (51). This study implemented a 3D variable density undersampled Cartesian trajectory. Additional navigation k-space data were periodically acquired to enable basis estimation for subspace construction. Meanwhile, this study aimed to reconstruct a six-dimensional image-set including a respiratory motion dimension, which could slow down the overall image reconstruction time. In comparison, our current study implements stack-of-stars sampling and aims to estimate a temporal basis from the centers of k-space without the need for sequence modification. Meanwhile, with stack-of-stars sampling, one can break down the reconstruction task for each slice separately, which can reduce the overall reconstruction burden and speed up the reconstruction process with parallel computing.

The performance of reconstructing an image series from only a few spokes per frame primarily relies on available temporal correlations. As such, our proposed method requires a sufficient number of dynamic frames. Figure 4 has shown that the expiratory image from GRASP-Pro reconstruction (with 5 spokes per dynamic frame) is comparable with XD-GRASP reconstruction (with 162 spokes per motion phase), while the inspiratory image from XD-GRASP reconstruction looks better than GRASP-Pro. This is mainly because the inspiratory duration in a normal breathing cycle is shorter than the expiratory duration. In addition, the construction of subspace also plays a key role in reconstructing highly-accelerated dynamic images. As previously demonstrated by Zhao et al (47), subspace-based reconstruction substantially reduces the degrees of freedoms compared to other low-rank reconstruction methods without generating a subspace (e.g., L+S), which explains why GRASP-Pro is superior to L+S as shown in Figure 9.

There are several advantages to apply the proposed 4D real-time MRI method to DCE-MRI, especially in the liver. First, DCE-MRI has to be performed in real time since contrast enhancement is not periodic. Second, DCE-MRI acquisition typically lasts for several minutes, which provides sufficient data to exploit temporal correlations. Third, our method provides a more robust way to address intra-frame respiratory blurring without the need for motion detection. Fourth, as shown in Figure 7, a large number of contrast phases can be reconstructed in DCE-MRI with our method, which offers additional flexibility (compared to standard GRASP with moderate temporal resolution) to retrospectively choose desired contrast phases. Fifth, it has been reported that the arterial phases in DCE-MRI of the liver are often degraded by respiratory motion due to the quick passage of a contrast agent (56). By addressing intra-frame respiratory blurring, our method could provide an easy way to address this challenge.

Our proposed 4D real-time MRI method aims to address intra-frame motion. However, with this target temporal resolution, inter-frame motion (displacement from one dynamic frame to the others caused by respiration) can become a new challenge in DCE-MRI of the liver. This is generally not a problem for qualitative evaluation in the clinic, as radiologists typically look at one specific contrast phase through all the slices to assess the enhancement of lesions. However, the existence of inter-frame motion can restrict perfusion quantification. An easy way to address this is to use only the expiratory phases for quantitative analysis, but the distribution of the phases along time may not be uniform since it is dependent on underlying breathing. Another approach to solve this problem is to perform another reconstruction on the expiratory phases to fill in the missing frames. This can be performed with simple interpolation or based on low-rank matrix completion, assuming that the underlying contrast enhancement is smooth. These hypotheses will be evaluated in future works.

The proposed 4D real-time MRI method can have more applications beyond those demonstrated in this work. It can be applied for imaging moving joints (e.g., wrist or knee) or for imaging speech with volumetric information. Image-guided treatment planning or adaptive radiotherapy is another nice application to use this imaging approach. Without the need of self-gating or motion sorting, additional robustness can be obtained to address potential motion drift in Image-guided treatment. One limitation of the method, however, is that the achieved temporal resolution can only address respiratory motion but not cardiac motion. This is because cardiac motion occurs at a much higher frequency than respiratory motion, which may need a temporal resolution of less than 100ms. As a result, the target application of this imaging method is limited to the body or chest but not the cardiovascular system.

## Conclusion

This study proposed a highly-accelerated 4D real-time MRI framework that combine 3D golden-angle radial stack-of-stars sampling and tailored GRASP-Pro reconstruction to enable sub-second temporal resolution. The ability to acquire a 3D image within one second intrinsically provides a simple but effective method to reduce intra-frame respiratory motion blurring for body applications and eliminates the need for motion detection and motion compensation. The method could potentially be useful for different applications that benefit from real-time imaging with free-breathing volumetric information, such as DCE-MRI and image-guided treatment planning.

## Acknowledgement

This work was supported in part by the NIH (R01EB030549, R01EB031083, R21EB032917). Early support of this work was derived from NIH P41 EB017183. The author thanks Ding Xia for help with the T1 mapping phantom experiment, and thank Dr. Wei-Ching Lo and Dr. Nicole Seiberlich for sharing the MATLAB code for the abdominal phantom simulation.

## Disclosure

Li Feng is a co-inventor of a patent (patent number 9921285) on the GRASP MRI technique.